\documentclass[conference]{IEEEtran}
\ifCLASSINFOpdf
\else
\fi
\usepackage{graphicx}
\usepackage{amsmath}
\usepackage{booktabs}
\usepackage{multirow}
\usepackage{hyperref}
\hypersetup{colorlinks=true}
\begin{document}
\setlength{\abovedisplayskip}{3pt}
\setlength{\belowdisplayskip}{3pt}

\title{Fast and Energy-Efficient CNN Inference\\ on IoT Devices}

\author{
\IEEEauthorblockN{Mohammad Motamedi}
\IEEEauthorblockA{ECE Department\\
University of California, Davis\\
Email: mmotamedi@ucdavis.edu}
\and
\IEEEauthorblockN{Daniel Fong}
\IEEEauthorblockA{ECE Department\\
	University of California, Davis\\
	Email: dfong@ucdavis.edu}
\and
\IEEEauthorblockN{Soheil Ghiasi}
\IEEEauthorblockA{ECE Department\\
University of California, Davis\\
Email: ghiasi@ucdavis.edu}
}

\maketitle

\begin{abstract}
Convolutional Neural Networks (CNNs) exhibit remarkable performance in various machine learning tasks. As sensor-equipped internet of things (IoT) devices permeate into every aspect of modern life, it is increasingly important to run CNN inference, a computationally intensive application, on resource constrained devices. We present a technique for fast and energy-efficient CNN inference on mobile SoC platforms, which are projected to be a major player in the IoT space. We propose techniques for efficient parallelization of CNN inference targeting mobile GPUs, and explore the underlying tradeoffs. Experiments with running Squeezenet on three different mobile devices confirm the effectiveness of our approach. For further study, please refer to the project repository available on our GitHub page: {\href{https://github.com/mtmd/Mobile_ConvNet}{https://github.com/mtmd/Mobile\_ConvNet}}.
\end{abstract}

\IEEEpeerreviewmaketitle

\section{Introduction}
Convolutional Neural Networks (CNNs) have been extensively used in different image and video processing applications \cite{iandola2016squeezenet,krizhevsky2012imagenet,szegedy2015going}. Even though CNNs have remarkable performance, they are computationally intensive. There have been different proposals for accelerating CNNs using GPUs (\cite{jia2014caffe,oskouei2015gpu}), FPGAs (\cite{zhang2015optimizing,Motamedi2016Design}) and ASICs. Among these platforms, GPUs are the main platform for accelerating Convolutional Neural Network due to their high performance and ease of use.\\
There are two classes of GPUs: server-class and mobile-class. Server-class GPUs have more computational capabilities and consume a considerable amount of energy. These GPUs are used in cloud-based computing for training CNNs and processing computations which are offloaded to a cloud from mobile devices. Accelerating CNNs on server-class GPUs is very well studied. Nowadays, many applications rely on cloud computing to take advantage of machine learning techniques with CNNs. However, cloud computing is not always the best solution for several reasons. First, using a wireless media for transferring data to a cloud requires a considerable amount of energy. Second, sometimes it is not possible to offload computations to a cloud in locations with weak signal reception. Finally, due to privacy concerns, using remote processing infrastructures is not always a solution. \\
 Unlike server GPUs, mobile GPUs have limited hardware resources and they are designed to work within restricted power budgets. In this paper, we have used RenderScript for accelerating CNNs on mobile GPUs. RenderScript is a framework for utilizing heterogeneous computing on Android phones. During execution, the runtime engine distributes the computation on available processing elements such as CPU cores and GPU.\\
In this paper, we propose a solution for utilizing RenderScript features to implement CNNs efficiently on Android devices. Moreover, we will explain that the design parameters which yield the minimum execution time vary between different platforms and layers of a CNN. Subsequently, we implement and optimize SqueezeNet \cite{iandola2016squeezenet} on three different mobile devices and find the optimal design parameters for each.\\
The approach we propose in this paper can be used for accelerating any CNN efficiently. Having an execution time of less than a quarter of a second and an energy consumption of half a joule, the proposed algorithm makes it feasible to locally use CNNs on mobile devices.
\begin{figure}
	\includegraphics[width=\columnwidth]{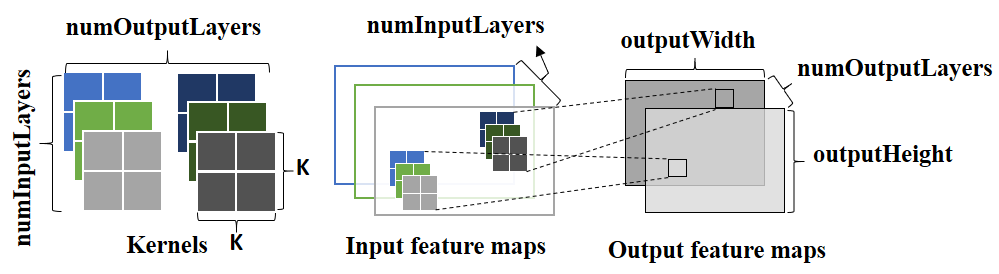}
	\centering
	\caption{\small Feature extraction with convolutional kernels.}
	\label{fig:conv}
\end{figure}
\section{Convolutional Neural Networks}
\label{DCNN}
Convolutional Neural Networks (CNNs) have millions of parameters, which are obtained during a training procedure.
In this work, we focus on the acceleration of the forward path using mobile GPUs. \\
Each CNN has multiple convolutional layers. Each layer uses 3D kernels
for feature extraction. The result of multiple feature extractions are added together to create Output Feature Maps. The convolution operation for two filter banks that each of them has three kernels is shown in Figure (\ref{fig:conv}).
The number
of input feature maps, output feature maps, and the output size are $numInputLayers$, $numOutputLayers$, and $outputWidth \times outputHeight$, respectively.\\
One kernel has dimension $K \times K$, and a layer has $numInputLayers \times numOutputLayers$ of these kernels.
Each pixel in an Output Feature Map is the sum of a convolutions between inputs and the respective kernels. To generate adjacent pixels in an Output Feature Map, the kernel bank is slid across the spacial dimension by stride $S$. The convolution operation is shown in Figure (\ref{fig:conv_code}).\\
In this paper, we use SqueezeNet \cite{iandola2016squeezenet} as a use case. SqueezeNet has four different types of layers: convolutional, pooling, softmax, and fire. The first three layers are commonly used in many CNN architectures. Fire layers are unique to SqueezeNet and consist of three convolutional layers: one squeeze layer and two expand layers. More than 90\% of the execution time in CNNs is spent in convolutional layers. \\
SqueezeNet has eight fire layers and two convolutional layers. The input to the very first layer is a 224 x 224 RGB image. Convolutional and fire layers are used to extract features from the input image. The last layer is a classifier with a thousand output nodes, each of which is a prediction for the different image categories.\\
\section{Parallel Algorithm Design}
\begin{figure}
	\includegraphics[width=7cm]{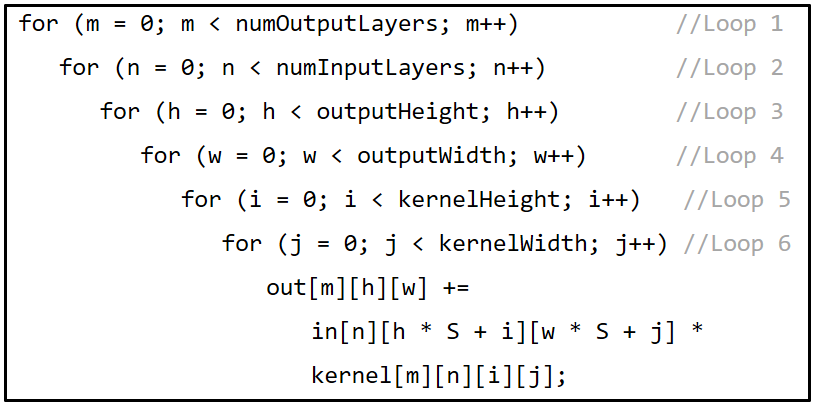}
	\centering
	\caption{Sequential Implementation of a convolution kernel.}
	\label{fig:conv_code}
\end{figure}
A sequential implementation of the convolution operation is shown in Figure (\ref{fig:conv_code}). Loops \#2 to \#6 perform a 3D convolution (both input and kernel are 3D data structures). Loop \#1, repeats the 3D convolution for different kernels in order to compute different layers of the output. Recall that each layer of the output is the result of a 3D convolution of the input with a distinct filter bank. 
In Figure (\ref{fig:conv_code}), S (stride) is the number of pixels that a kernal slides during each step of the convolution.\\
The convolution operation takes a major part of the execution time in CNNs. In this section, we propose an approach for accelerating the convolution using RenderScript on mobile devices.
\begin{equation}
\small\begin{split}
numOutputElements = numOutputLayer \times outputHeight \\ \times outputWidth
\end{split}
\label{EQ: numOutputelems}
\end{equation}
\subsection{Parallel Computation of Output Elements}
\label{sec: ParaCompute}
\begin{figure}
	\includegraphics[width=7cm]{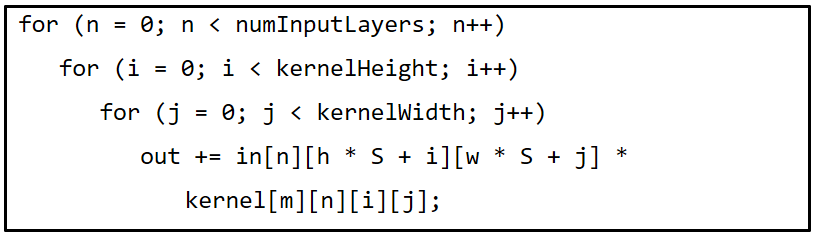}
	\centering
	\caption{Each thread of the output allocation runs this code for a distinct part of the input to generate an output element. Values of $w, h$, and $m$ are computed using Equations (\ref{EQ: w}), (\ref{EQ: h}), and (\ref{EQ: m}), respectively.}
	\label{fig:code1}
\end{figure}
Output of each convolution layer is a 3D matrix which includes $numOutputElements$ elements (Equation (\ref{EQ: numOutputelems})).
Computing these elements is a perfectly parallel workload. In theory, it is possible to assign one thread per element and execute all these threads concurrently. To implement this in RenderScript, it is required to define an allocation\footnote{Allocation is a class in RenderScript which is designed for passing data to a RenderScript kernel. More information is available on Android website:\\ https://developer.android.com/reference/android/renderscript/Allocation.html} with dimensions matching the size of the output matrix. Successively, for each element of that allocation, it is required to compute the result of a dot product between the corresponding window of the input and a specific kernel. This dot product is performed using the pseudo code shown in Figure (\ref{fig:code1}). Notice that this code has to be executed for every element of the output allocation. In each allocation, every thread has an index $x$. For the aforementioned output allocation $x \in [0, numOutputElements - 1]$. In each thread, index $x$ is used to determine values of $w$, $h$, and $m$ as it is shown in Equations (\ref{EQ: w}), (\ref{EQ: h}), and (\ref{EQ: m}) respectively. The value of $(m, w, h)$ in each thread is unique. These parameters help a thread to work on a distinct part of inputs and generate a specific element in the output matrix. 
\begin{equation}
w = x~\%~outputWidth
\label{EQ: w}
\end{equation}
\begin{equation}
h = \lfloor\frac{x}{outputWidth}\rfloor~\%~outputHeight
\label{EQ: h}
\end{equation}
\begin{equation}
m = \lfloor\frac{x}{outputWidth \times outputHeight}\rfloor
\label{EQ: m}
\end{equation}
\subsection{Utilizing Vectorized Operations}
\begin{figure}
	\includegraphics[width=7cm]{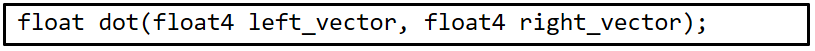}
	\centering
	\caption{A prototype of vectorized dot function which is offered in RenderScript. The RenderScript type \texttt{float4} is used to represent a vector of four elements of type float. The input of this function are two vectors of type float and the output is the dot product of these two vectors.}
	\label{fig:dot}
\end{figure}
RenderScript offers various vector math functions that interpret their input arguments as a representation of vectors. Currently, the maximum supported length for these vectors is four. These built-in functions can be efficiently mapped on supported GPUs. A prototype of the vectorized dot function is shown in Figure (\ref{fig:dot}). In this case, inputs are two vectors, each of which includes four floating point numbers. The function performs a dot operation between the two vectors and returns the result as a floating point number. \\
We use the vectorized dot function for further acceleration of the convolution. To do so, it is required to change the implementation offered in the pseudo code of Figure (\ref{fig:code1}). As previously discussed, this code computes the dot product between a certain window of the input and a kernel. This process is performed in parallel for all elements of the output allocation. However, the process itself is sequential (three nested loops in Figure (\ref{fig:code1})). In the following subsections, we explain how to accelerate it using the vectorized dot function offered in RenderScript.
\subsubsection{Data Reordering}
\label{Data Reordering}
\begin{figure}
	\includegraphics[width=7cm]{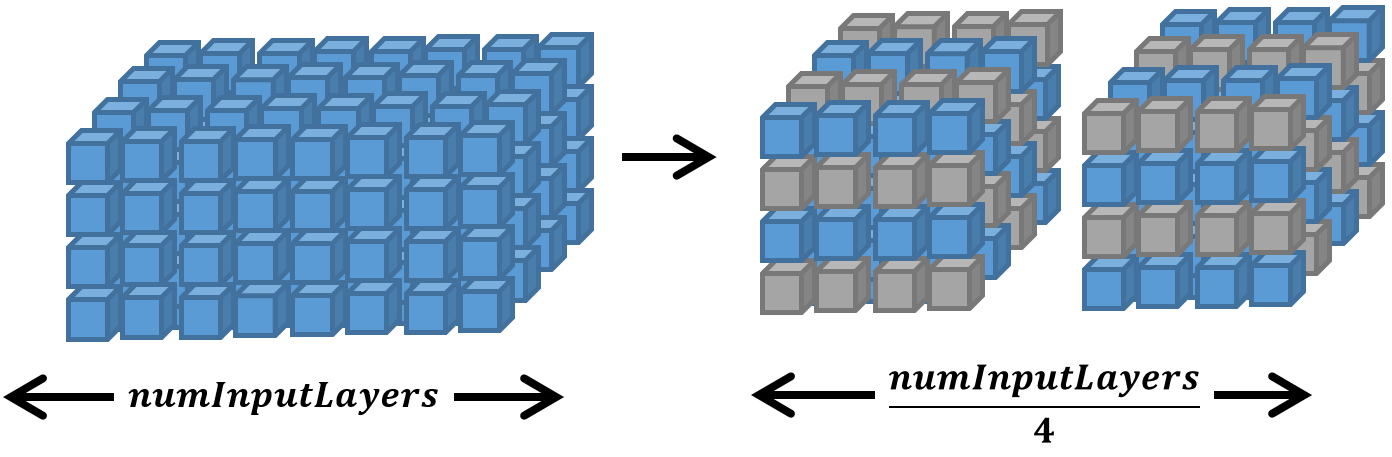}
	\centering
	\caption{It is required to change the data structure from traditional row or column major to a vectorized form. Elements in the same spacial location from consecutive layers form a vector. Each four elements in gray or blue form a vector.}
	\label{fig:reshape}
\end{figure}

By default, both input and convolution kernels are stored in either row or column major order. Therefore, elements stored in the vicinity of each other are either the next element from the same row (column) or the first element of the next row (column). If we represent each element using $(Layer, Row, Column)$, then in a row major fashion, data is stored in the following format:\\
\begin{equation}
\tiny D = \{(0, 0, 0), (0, 0, 1), (0, 0, 2) \cdots, (0, 1, 0), (0, 1, 1), \cdots, (0, 2, 0), (0, 2, 1), \cdots\}
\label{EQ: row}
\end{equation}
To utilize the vectorized dot product, we need to make small vectors with a length of four. Hence, it is required to change the previous data representation to the one shown in Equation (\ref{EQ: vectorized}). A 3D representation of this transform is shown in Figure (\ref{fig:reshape}). 
\begin{equation}
\label{EQ: vectorized}
\tiny\begin{split}
D^{'} = \{(0, 0, 0), (1, 0, 0), (2, 0, 0), (3, 0, 0), (0, 0, 1), (1, 0, 1), (2, 0, 1), (3, 0, 1), \cdots, \\(4, 0, 0), (5, 0, 0), (6, 0, 0), (7, 0, 0), \cdots\}
\end{split}
\end{equation}
\subsubsection{Vectorized dot Function} 
\begin{figure}
	\includegraphics[width=7cm]{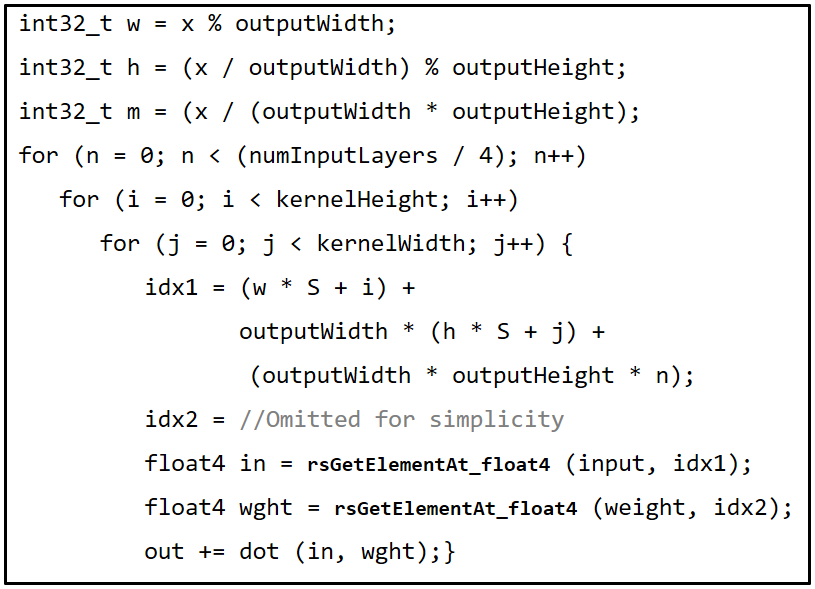}
	\centering
	\caption{Vectorized implementation of the pseudo code of Figure (\ref{fig:code1}). Function \texttt{rsGetElement\_float4} reads a vector from memory and \texttt{dot} performs a vectorized dot product.}
	\label{fig:vectorized_code}
\end{figure}

When the data is reordered, it is sufficient to read them as vectors with a length of four, perform a vectorized dot product, and accumulate the result. This is implemented in the pseudo code shown in Figure (\ref{fig:vectorized_code}). The function \small{\texttt{rsGetElementAt\_float4}} reads four floating point numbers from the input vector. The function has two inputs, the first input is the pointer to the memory location from which elements should be read and the next input is the base address. \\
Using this function, we read four different values from four consecutive layers and process them at the same time. In each iteration of the outermost loop, four different layers are processed. This reduces the total number of iterations by a factor of four when compared to the pseudo code in Figure (\ref{fig:code1})).
\subsection{Zero-overhead Vectorization}
\begin{figure}
	\includegraphics[width=5cm]{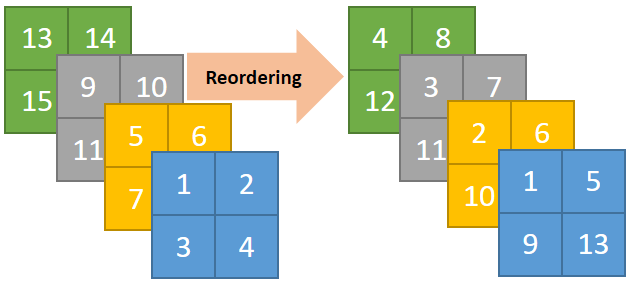}
	\centering
	\caption{Before reordering, data is stored in the row major format. After reordering, data is stored in a layer major format.}
	\label{fig:data_reshape2}
\end{figure}
As we explained in section (\ref{Data Reordering}), in order to use vectorized operations, it is required to change the order of elements in the input and convolution kernels. Changing the order in convolution kernels can be performed offline. They can be reordered once, reshaped, and rewritten in a new model file. However, reordering the input is complicated. In convolutional neural networks, the output of each layer is the input to the next layer. Therefore, it is required to reorder the output of the previous layer before each convolution. This process is time and energy consuming. In this section, we will explain how to generate the output in a vectorized form to avoid overheads of data reordering.\\
As we explained in section (\ref{sec: ParaCompute}), each thread has a unique index $x$. We offered Equations (\ref{EQ: w}), (\ref{EQ: h}), and (\ref{EQ: m}) for computing $w$, $h$, and $m$, respectively. These parameters determine the location of the output element that thread $x$ generates. Therefore, in order to generate the output of the convolution in a reordered format, it is required to change these parameters. For example, the result of computations in the second thread ($x = 1$) is stored in the second location of the output memory. After reordering, however, the second element of the output array should be $(m = 1, w = 0, h = 0)$. This is illustrated in Figure (\ref{fig:data_reshape2}).\\
To create the output in the reordered format, it is required to generate indexes for stacks of four layers, instead of a single layer (Figure (\ref{fig:data_reshape2})). Hence, we start indexing the second row when we have indexed all first rows of all four layers. Therefore, $w$ and $h$ can be computed using Equations (\ref{EQ: w_new}) and (\ref{EQ: h_new}), respectively. For computing the value of $m$ (layer index), it is required to see which stack and layer a particular output belongs to (Figures (\ref{fig:reshape}) and Figure (\ref{fig:data_reshape2})). Equation (\ref{EQ: m_new}) computes the value of $m$.
\begin{equation}
w = \lfloor x / 4\rfloor~\%~outputWidth
\label{EQ: w_new}
\end{equation}
\begin{equation}
h = \lfloor\frac{x}{4 \times outputWidth}\rfloor~\%~outputHeight
\label{EQ: h_new}
\end{equation}
\begin{equation}
m = (x~\%~4) + \lfloor\frac{x}{4 \times outputWidth \times outputHeight}\rfloor \times 4
\label{EQ: m_new}
\end{equation}
\begin{figure}
	\includegraphics[width=7cm]{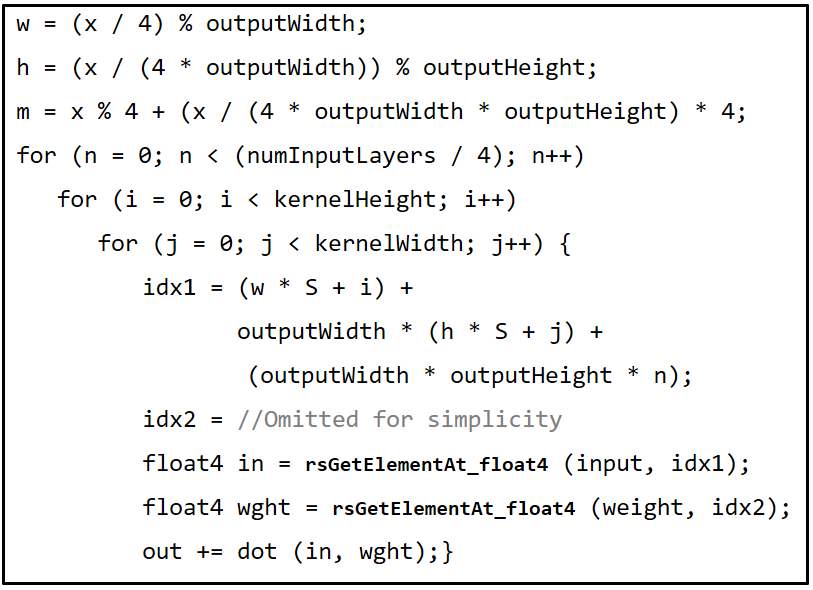}
	\centering
	\caption{Reimplementation of pseudo code of Figure (\ref{fig:vectorized_code}) using zero overhead vectorization approach.}
	\label{fig:fullVec}
\end{figure}
Using Equations (\ref{EQ: w_new}), (\ref{EQ: h_new}), and (\ref{EQ: m_new}), it is possible to generate the output in the reordered format. Such an output can be directly used as the input to the next layer without additional processing overhead. The pseudo code in Figure (\ref{fig:fullVec}) shows the final implementation. 
\subsection{Optimizing Thread Granularity}
\label{sec: Optimizing Thread Granularity}
\begin{figure}
	\includegraphics[width=7cm]{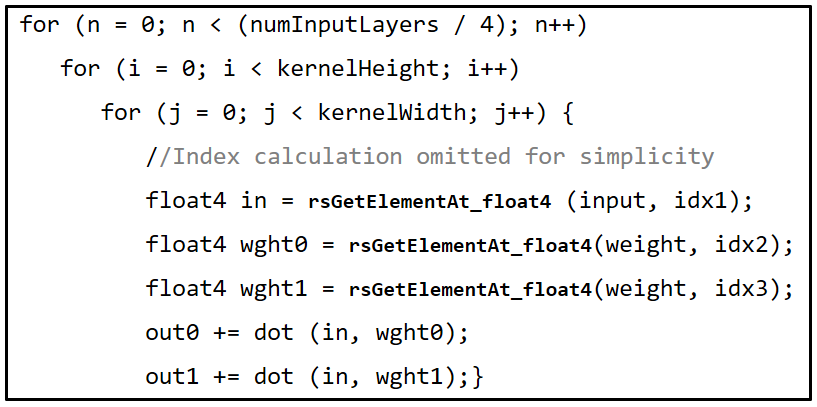}
	\centering
	\caption{Each thread computes two output elements: \texttt{out0} and \texttt{out1}.}
	\label{fig:fullVecGra}
\end{figure}
There is a trade-off between the number of threads and the amount of code that should be executed in each thread. Decreasing the number of threads increases the amount of code per thread (i.e., each thread has to compute multiple elements of output instead of one). Increasing the number of threads beyond some hardware limit will not improve the execution time since there is not sufficient resources for parallel execution of all of these threads. In addition, when a thread is responsible for generating more than one output value, it is possible to decrease memory access by data reusability. Therefore, in this problem, defining the highest possible number of threads is not the optimal solution. In the experimental results we show that the shortest execution time does not belong to the finest thread granularity (highest number of threads).\\
In order to find the optimal thread granularity, we have implemented convolution kernels with dynamic thread granularities. In this paper, we use the keyword \texttt{conv\_g} for referring to them. Where, \texttt{g} is the number of output elements that threads compute sequentially. For example, kernel \texttt{conv\_4} computes four elements sequentially. For smaller values of \texttt{g} thread granularity is finer (larger number of threads that each of them has a smaller task). In Figure (\ref{fig:fullVec}) the value of \texttt{g} is 1 and Figure (\ref{fig:fullVecGra}) shows the implementation for \texttt{g = 2}. When \texttt{g} is larger than one, the input values are loaded to thread memory once, but are used \texttt{g} times. Moreover, when the value of \texttt{g} is larger than one, a thread has to compute the value of an element $(i, j)$ for more than one output layers. For example, when \texttt{g = 2}, the same thread that computes the value of element $(i, j)$ in layer 0, has to compute the value of element $(i, j)$ in layer $numOutputLayers / g$. However, as we described, output layers are generated in the form of vectors of length four. Therefore, when changing the thread granularity, it is important to make sure that $numOutputLayers / g$ is divisible by four. 
\subsection{Pooling and Softmax}
There are two types of pooling functions in SqueezeNet: max pooling and average pooling. As their names explain, the former finds the maximum and the latter computes the average of numbers in a given window. We have used vectorized function \texttt{fmax} and \texttt{sum} for RenderScript based implementation of these functions. Implementations are analogous to convolution layers. The execution time of the softmax function is negligible; hence, a GPU based implementation of this function is not required. \\
In this section we explained the approach used for RenderScript based implementation of SqueezeNet. For simplicity, we skipped a considerable amount of details. For further study, please refer to the project repository available on our GitHub page\footnote{\href{https://github.com/mtmd/Mobile_ConvNet}{https://github.com/mtmd/Mobile\_ConvNet}}.
\section{Experimental Results}
\begin{table*}
	\centering
	\caption{Optimal thread granularities for SqueezeNet on different platforms.}
	\scalebox{0.9} {
		\begin{tabular}{lccccccccccccc}
			\toprule[.05cm]
			& Conv1 & F2EX1 & F2EX3 & F3EX1 & F3EX3 & F4EX1 & F4EX3 & F5EX1 & F5EX3 & F6EX1 & F6EX3 & F7EX1 & F7EX3 \\ \toprule[.05cm]
			Galaxy S7 &  G6   &  G8   &  G4   &  G8   &  G8   &  G8   &  G8   &  G4   &  G4   &  G12  &  G12  &  G6   &  G4   \\ \midrule
			Nexus 6P  &  G6   &  G8   &  G4   &  G8   &  G4   &  G8   &  G4   &  G8   &  G4   &  G16  &  G6   &  G6   &  G6   \\ \midrule
			Nexus 5   &  G12  &  G8   &  G16  &  G8   &  G16  &  G8   &  G8   &  G32  &  G8   &  G12  &  G12  &  G12  &  G12  \\ \bottomrule[.05cm]
		\end{tabular}
	}
	\label{TBL:optimalGra}	
\end{table*}
\begin{figure*}
	\includegraphics[width=\textwidth]{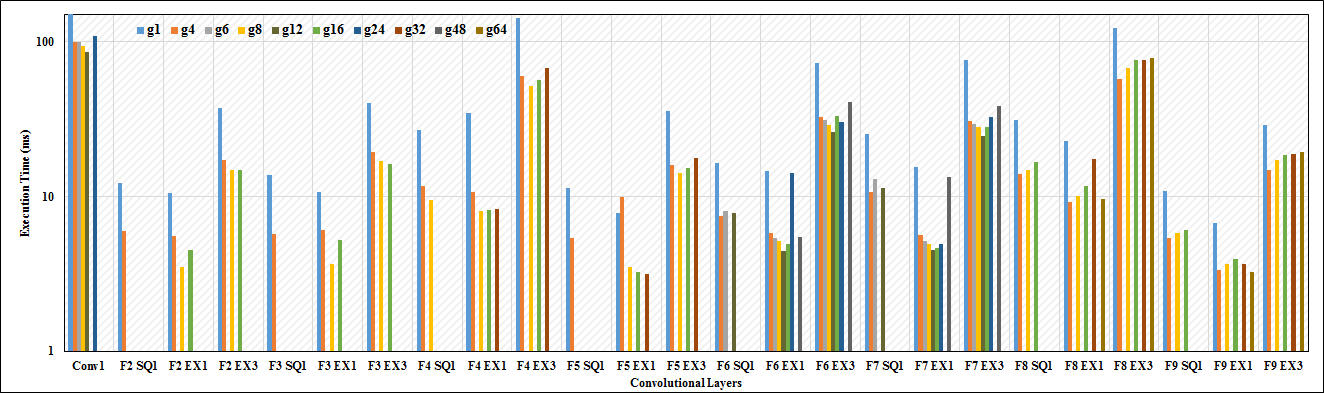}
	\centering
	\caption{Execution time of layers of SqueezeNet for different thread granularities. Numbers are measured on Nexus 5. Highest number of threads (\texttt{g = 1}) has the worst execution time.}
	\label{fig:geffect}
\end{figure*}
\begin{table}
	\centering
	\caption{Hardware specifications of mobile devices which are used in evaluations.}
	\scalebox{0.7} {
	\begin{tabular}{lcccc}
		\toprule[.05cm]
		                  &      SoC       & Tech. &  CPU Frequency   &         GPU         \\ \toprule[.05cm]
		Samsung Galaxy S7 & Snapdragon 820 & 14 nm & 0.307 - 2.15 GHz & Adreno 530 @624 MHz \\ \midrule
		Huawei Nexus 6P   & Snapdragon 810 & 20 nm & 0.384 - 1.96 GHz & Adreno 430 @650 MHz \\ \midrule
		LG Nexus 5        & Snapdragon 800 & 28 nm &  0.3 - 2.27 GHz  & Adreno 330 @450 MHz \\ \bottomrule[.05cm]
	\end{tabular}
}
	\label{TBL:Platforms}	
\end{table}	
\begin{table}
	\caption{Effect of thread granularity. Choosing optimal granularity over pessimal improves the execution time by at least 2X.}
	\label{TBL: granularity}
	\centering
	\scalebox{0.7}{
	\begin{tabular}{lccccccc}
		\toprule[.05cm]
		                                                  & \multicolumn{3}{c}{Fire Layers (ms)} & \multicolumn{3}{c}{Convolutional Layers (ms)} & \multirow{2}{*}{Speedup} \\
		\cmidrule[.05cm](l){2-4} \cmidrule[.05cm](l){5-7} & {Optimal} & {Pessimal} &  {Speedup}  & {Optimal} & {Pessimal} &     {Speedup}      &  \\ \midrule
		Galaxy S7                                         &  268.94   &   852.36   &   3.17X    &  159.55   &   227.46   &       1.43X       &         2.52X         \\ \midrule
		Nexus 6P                                          &  234.72   &   543.01   &   2.31X    &  134.91   &   205.19   &       1.52X       &          2.02X          \\ \midrule
		Nexus 5                                           &   323.6   &   828.35   &   2.56X    &  247.59   &   475.84   &       1.92X       &          2.28X          \\ \bottomrule[.05cm]
	\end{tabular}
}
\end{table} 
We have used three different mobile devices for evaluating the proposed RenderScript based parallel algorithm. The specifications of these phones are shown in Table (\ref{TBL:Platforms}). These phones are equipped with three major Qualcomm Snapdragon SoCs. \\
In order to increase the precision, all experiments of this section have been repeated 10 times, the average is computed and reported. During the experiments the phone's background processes were stopped, placed in airplane mode, and fully-dimmed their screen brightness. In this section, we present the results of executing the proposed algorithm on these phones and perform some analysis.\\
We have accelerated SqueezeNet\cite{iandola2016squeezenet} using the proposed approach. SqueezeNet has two normal convolutional layers and nine fire layers. Each fire layer has three convolutional layers: two squeeze layers and one expand layer. We use \texttt{Fn SQn} and \texttt{Fn EXn} notations to refer to squeeze and expand layers, respectively.
\subsection{The Effect of Thread Granularity}
As we explained in section (\ref{sec: Optimizing Thread Granularity}), for each convolutional layer in SqueezeNet, there is a finite set of valid values for \texttt{g}. The result of implementation of SqueezeNet for different values of this set is shown in Figure (\ref{fig:geffect}). \\
As Figure (\ref{fig:geffect}) illustrates, having the finest thread granularity (\texttt{g = 1}) is not the optimal solution for any layer. By increasing the value of \texttt{g}, data reusability increases, thereby reducing the execution time. However, after some threshold, any further increase of the value of \texttt{g} exacerbates the execution time. Notice that implementations based on very large values of \texttt{g} do not use the available parallel resources efficiently. \\
Thread granularity plays an important role in determining the execution time. In Table (\ref{TBL: granularity}) we show the execution time of SqueezeNet for three different smart phones. For all of these phones, using optimal thread granularity drastically decreases the execution time. Even on modern platforms, such as Galaxy S7, optimal thread granularity can yield a speedup of 2.52X compared to a pessimal thread granularity. \\
Finally, the optimal solution for thread granularity for some layers of SqueezeNet are shown in Table (\ref{TBL:optimalGra}). For some layers, such as F2 EX1, F3 EX1, and F4 EX1, the optimal solution is identical for different platforms. However, the optimal thread granularity varies based on the convolution layer specifications and the target hardware. 
\subsection{Imprecise Computing}
	\begin{table*}
		\centering
		\caption{Execution time (millisecond) of layers of SqueezeNet using sequential, parallel and imprecise parallel algorithms.}
		\scalebox{0.9} {
			\begin{tabular}{llcccccccccc}
				\toprule[.06cm]
				                          & Algorithm          & Conv 1  & Fire 2  & Fire 3  & Fire 4  & Fire 5  & Fire 6  & Fire 7  & Fire 8  & Fire 9  & Conv 10 \\ \toprule[.05cm]
				                          & Sequential         & 2523.64 & 526.36  & 561.73  & 2029.00 & 490.09  & 1044.27 & 1138.00 & 1929.91 & 428.91  & 1567.91 \\
				\cmidrule{2-12} Galaxy S7 & Precise Parallel   &  55.82  &  25.55  &  26.00  &  46.37  &  20.82  &  33.91  &  33.92  &  58.55  &  23.82  & 103.73  \\
				\cmidrule{2-12}           & Imprecise Parallel &  7.10   &  18.10  &  15.60  &  21.50  &  20.70  &  20.10  &  23.10  &  22.70  &  17.80  &  19.00  \\ \midrule[.05cm]
				                          & Sequential         & 3275.55 & 711.82  & 743.82  & 2652.18 & 678.45  & 1422.82 & 1530.27 & 2606.73 & 617.00  & 2954.09 \\
				\cmidrule{2-12} Nexus 6P  & Precise Parallel   &  52.36  &  18.54  &  19.45  &  44.82  &  17.82  &  29.00  &  27.91  &  55.73  &  21.45  &  82.55  \\
				\cmidrule{2-12}           & Imprecise Parallel &  12.40  &  9.87   &  8.89   &  12.07  &  7.55   &  9.11   &  10.48  &  16.04  &  13.78  &  15.40  \\ \midrule[.05cm]
				                          & Sequential         & 8873.45 & 1835.91 & 2015.73 & 7232.36 & 1889.55 & 3983.09 & 4285.73 & 7220.91 & 1571.18 & 4742.73 \\
				\cmidrule{2-12} Nexus 5   & Precise Parallel   &  86.17  &  24.33  &  25.59  &  69.00  &  22.76  &  37.92  &  39.83  &  80.67  &  23.50  & 161.42  \\
				\cmidrule{2-12}           & Imprecise Parallel &  16.70  &  13.30  &  9.20   &  17.50  &  8.50   &  9.10   &  9.20   &  12.60  &  8.40   &  26.40  \\ \bottomrule[.05cm]
			\end{tabular}
		}
		\label{TBL:finalResults}	
	\end{table*}
\begin{table*}
	\caption{Power and energy consumption of SqueezeNet using sequential and parallel algorithms.}
	\label{TBL: power}
	\centering
	\scalebox{0.9}{
		\begin{tabular}{lccccccccccccc}
			\toprule[.05cm]
			&       \multicolumn{3}{c}{Total Power (mW)}       & \multicolumn{2}{c}{Differential Power (mW)} & \multicolumn{2}{c}{Energy/Image (J)} & \multirow{2}{*}{Energy Ratio} &  \\\cmidrule[.04cm](l){2-4} \cmidrule[.04cm](l){5-6} \cmidrule[.04cm](l){7-8} 
			& {Baseline} & {Sequential} & {Imprecise Parallel} & {Sequential} &     {Imprecise Parallel}     & {Sequential} & {Imprecise Parallel}  &  \\ \midrule
			Galaxy S7 &   173.18   &   1552.51    &       2921.79        &   1379.33    &           2748.61            &      17      &         0.569         &            \textbf{29.88X}             &  \\ \midrule
			Nexus 6P  &  1480.97   &   1999.12    &       5461.89        &    518.15    &           3980.92            &     8.96     &         0.514         &            \textbf{17.43X}             &  \\ \midrule
			Nexus 5   &   422.71   &    1023.3    &       1170.45        &    600.29    &            747.74            &    26.37     &         0.106         &            \textbf{249.47X}            &  \\ \bottomrule[.05cm]
		\end{tabular}
	}
\end{table*} 	
RenderScript offers two modes for imprecise computing for applications that do not need a strict implementation of IEEE 754. These modes are called relaxed and imprecise floating point computation. The relaxed floating point mode enables flush to zero for denormalized numbers and round toward zero. The imprecise computing mode enables everything in the relaxed computing. In addition, in this mode operations resulting in -0.0 can return +0.0 and operations on INF and NAN are undefined. \\
Using imprecise computing accelerates the execution time since some optimizations are only available for computation with relax/imprecise precisions. Notice that relaxed and imprecise are GPU modes and using these modes does not have any effect on the CPU side.\\
 We have implemented SqueezeNet with both relaxed and imprecise precisions. Subsequently, we have tested the implementation on the first 10000 samples of ILSVRC 2012 validation dataset \cite{ILSVRC15}. For all these samples, the prediction results are identical to the original predictions of SqueezeNet. Therefore, using either of these modes does not change the classification accuracy. \\
Table (\ref{TBL:finalResults}) includes the execution time of different convolutional layers of SqueezeNet. Execution time was measured on three different phones. For each platform, the first row shows execution times for sequential execution. Second row includes times for parallel execution, and the third row shows execution times for imprecise parallel implementation. The units of all execution times is in milliseconds.\\
Table (\ref{TBL:speedup}) shows total execution time (all layers including pooling and softmax) of sequential and parallel implementation of SqueezeNet. In precise computing mode, processing each image takes between 388.36 (ms) on Nexus 6P  to 588.29 (ms) on Nexus 5. As Table (\ref{TBL:speedup}) shows, the parallel algorithm is drastically faster than the sequential implementation of SqueezeNet on each platform. In precise processing mode, the proposed parallel algorithm is at least 28.24X (Galaxy S7) and at most 74.68X (Nexus 5) faster than the sequential equivalent. \\
As we explained, using imprecise computing does not have any impact on the accuracy of SqueezeNet. However, using imprecise computing decreases the execution time drastically by using SIMD optimization of GPUs. On Galaxy S7, the imprecise parallel algorithm is 2.11X faster than the precise parallel implementation. On Nexus 5 and Nexus 6P, imprecise parallel algorithms are 4.16X and 3X faster than precise parallel implementation, respectively. Using imprecise computing, the total execution time of SqueezeNet varies between 129.21 millisecond (Nexus 6P) to 207.1 millisecond (Galaxy S7). The speedup of imprecise implementation of proposed parallel algorithm is 59.54X (Galaxy S7), 133.89X (Nexus 6P), and 310.74X (Nexus 5) compared to the basic sequential implementation. \\
Speedup factors which are shown in Table (\ref{TBL:speedup}) demonstrate that efficient use of mobile GPUs can decrease the execution time of neural networks drastically and make it possible to locally utilize CNNs on mobile devices. 
\begin{table}
	\centering
	\caption{Total execution time (millisecond) of SqueezeNet on different platforms.}
	\scalebox{0.8} {
		\begin{tabular}{lccccc}
			\toprule[.05cm]
			          & Sequential & Precise Parallel & Speedup & Imprecise Parallel &     Speedup      \\ \toprule[.05cm]
			Galaxy S7 &  12331.82  &      436.71      & 28.24X  &       207.1        &      59.54X      \\ \midrule
			Nexus 6P  &  17299.55  & \textbf{388.36}  & 44.55X  &  \textbf{129.21}   &     133.89X      \\ \midrule
			Nexus 5   &  43932.73  &      588.29      & 74.68X  &       141.38       & \textbf{310.74X} \\ \bottomrule[.05cm]
		\end{tabular}
	}
	\label{TBL:speedup}	
\end{table}
\subsection{Power Consumption}
Trepn profiler is an application made by Qualcomm for performance monitoring of applications on mobile devices. Trepn especially performs well on mobile devices with Snapdragon processors. We have used Android Intents to call Trepn for automated performance monitoring of parallel and sequential implementation of SqueezeNet. In all experiments, mobile devices have been in the airplane mode and screen brightness is set to minimum. Moreover, in all experiments background processes of different applications have been stopped. \\
Table (\ref{TBL: power}) shows the result of performance monitoring using Trepn profiler. Baseline indicates the power consumption of the system in the idle state. For example, for Nexus 6P  the base  power consumption is 1480.97 milliwatts. Differential power shows the required power for running an algorithm. On Galaxy S7 and Nexus 6P the required power for parallel algorithm is more than a sequential algorithm, since the parallel implementation requires more active cores simultaneously. However, in Nexus 5, the required power in parallel is less than sequential. There are two reasons for this low power consumption. First, maximum clock frequency of the GPU in Nexus 5 is 200 MHz less than the two other platforms. Second, Nexus 5 is equipped with an older Snapdragon chipset which has a lower performance. For a task such as running SqueezeNet the difference in performance is not considerable. However, in applications with intense process requirement such as games, the two other platforms outperform Nexus 5. \\
The main performance indicator is energy consumption. Sequential algorithms might have a better power consumption, but since the execution time is longer than parallel algorithms, the total energy consumption is much larger. In the third column of Table (\ref{TBL: power}) total energy consumptions for both sequential and parallel algorithms are computed. Using the parallel algorithm, the required energy for processing a single image on different platforms varies between 0.106 to 0.569 joules.
\section{Conclusion}
In this paper we proposed a solution for accelerating CNNs on mobile GPUs. Using this approach we have accelerated SqueezeNet and ran it on three different mobile devices with different chipsets. Experimental results on different platforms show that the proposed algorithm achieves a speedup of at least 59.54X and at most 310.74X. In addition, the energy consumption of the proposed parallel algorithm is at least 29.88X and at most 249.47X less than the energy consumption of the sequential algorithm. Offering an execution time of less than a quarter of a second and around half a joule energy consumption, the proposed algorithm makes it feasible to use CNNs on mobile devices.
\section*{Acknowledgment}
The authors would like to thank NVIDIA corporation for donating GPUs which were used in this research. 
\bibliographystyle{abbrv}
{
	\bibliography{sigproc}}


\end{document}